%% file: prd2000.tex
\documentstyle[aps,epsfig]{revtex}
\begin{document}
{\flushright
  Dubna  preprint  
    JINR,  E2-99-261,
    hep-ph/9910389}
\phantom{.}

\begin{center}
\large{
A MULTILOOP IMPROVEMENT OF NON-SINGLET QCD EVOLUTION EQUATIONS
}\\[0.5cm]

S.~V.~Mikhailov
\footnote{E-mail: mikhs@thsun1.jinr.dubna.su}\\[0.5cm]

{\it Joint Institute for Nuclear Research,
Bogoliubov Laboratory of Theoretical Physics,\\
141980, Moscow Region, Dubna, Russia}\\[0.5cm]

\end{center}
\begin{abstract}
     An approach is  elaborated for calculation of ``all loop" contributions
     to the non-singlet evolution kernels from the diagrams with  renormalon
     chain insertions. Closed expressions are
     obtained for sums of contributions to kernels $P(z)$ for the DGLAP
     equation and $V(x,y)$ for the ``nonforward" ER-BL equation from these
     diagrams that dominate for a large value of $b_0$,
     the first $\beta$-function coefficient. Calculations
     are performed in the covariant $\xi$-gauge in a MS-like scheme. It is
     established that a special choice of the gauge parameter $\xi=-3$
     generalizes the standard ``naive nonabelianization" approximation.
     The solutions are obtained to the ER-BL evolution equation
     (taken at the ``all loop" improved kernel), which are in form similar
     to one-loop solutions. A consequence for QCD descriptions
     of hard processes and the benefits and incompleteness of the approach
     are briefly discussed.
\end{abstract}

\vspace {0.5cm}

\section{Introduction}
Evolution kernels are main ingredients of the well-known evolution
equations for the parton distribution of DIS processes  \cite{L75}
and for parton wave functions  \cite{BL80} in hard exclusive
reactions. These equations describe the dependence of parton
distribution functions and parton wave functions on the
renormalization parameter $\mu^2$. The calculations performed
beyond the one-loop approximation for the forward DGLAP evolution
kernel $P(z)$ \cite{CFP80,FLK81}, and what is more, for the
nonforward Efremov-Radyushkin--Brodsky-Lepage (ER-BL) kernel
$V(x,y)$ \cite{DR84,MR85} were challenged and complicated
technical tasks. 15 years later, the 3-loop results for these
kernels are not known yet, except for the first few elements of
anomalous dimension in DIS, obtained numerically in \cite{LRV94}.
In this situation, it seems useful to try other ways to gain
knowledge about high-order corrections to these kernels and to the
solutions to the corresponding equations.

 Here I  discuss the results of the diagrammatic analysis and
multiloop calculations of the DGLAP kernel $P(z)$ and ER-BL kernel
$V(x,y)$ in a certain class of the ``all-order" approximation of
perturbative QCD (pQCD). The corresponding diagrams include the
chains of one-loop self-energy parts (renormalon chains) into the
one-loop diagrams (see Fig. 1). The regular method of calculation
and resummation of the indicated classes of  diagrams for these
kernels based upon their simple forest structure has been
suggested in ~\cite{M97}. There was established that the resulting
series possesses a nonzero convergent radius, therefore the
infrared renormalons are absent in the kernels. The results of
that summation for both the kinds of kernels (DGLAP and ER-BL)
obtained earlier in the framework of a scalar model in six
dimensions with the Lagrangian
 $\displaystyle L_{int} = g\sum^{N_f}_i \left( \psi^{*}_i \psi_i \varphi \right)
_{(6)}$ with $N_f$ of the scalar ``quark" flavours ($\psi_i$) and
``gluon" ($\varphi$) are analyzed here for non-singlet QCD
kernels. For the readers convenience some important results of the
paper ~\cite{M97} would be recalled.

The insertion of the chain into the ``gluon" line (``chain-1" in
\cite{M97}) of the diagram in Fig.1 a,b and resummation over all
bubbles transforms the one-loop kernel $aP_0(z)= a\bar{z} \equiv
a(1-z)$ into the ``improved" kernel $P^{(1)}(z; A)$ \begin{eqnarray}\label{IntA}
\displaystyle aP_0(z)= a\bar{z} \stackrel{chain-1}{\longrightarrow} P^{(1)}(z; A)
= a\bar{z} \left[ (z)^{-A}(1-A)
\frac{\gamma_{\varphi}(0)}{\gamma_{\varphi}(A)} \right];
~\mbox{where}~A=a N_f \gamma_{\varphi}(0),
~a=\frac{g^2}{(4\pi)^{3}}. \end{eqnarray}Here, $\gamma_{
\psi(\varphi)}(\varepsilon)$ are one-loop coefficients of the
anomalous dimensions of quark (gluon at $N_f=1$) fields in
D-dimension ($D=6-2\varepsilon$) discussed in \cite{M97}; for the
scalar model $\gamma_{\psi}(\varepsilon)=
\gamma_{\varphi}(\varepsilon)=B(2-\varepsilon, 2-\varepsilon)
C(\varepsilon)$, and $C(\varepsilon)$ is a scheme-dependent factor
($C(0)=1$) corresponding to a certain choice of an $\overline{\rm MS}$--like
scheme. The argument $A$ of the function $\gamma_{\varphi}(A)$ in
(\ref{IntA}) is a standard anomalous dimension (AD) of a ``gluon"
field. On the other hand, the result (\ref{IntA}) corresponds to
resummation of a class of series\footnote{On the other hand these class simply corresponds
to the Taylor expansion of the kernel $P^{(1)}(z; A)$ in a new parameter $A$,
so, the $n$-term of expansion corresponds to the $n$-bubble chain insertion.}
 like
$\displaystyle a \frac{(-A)^n}{n!} \left(\ln[z]+8/3 \right)^n$-series,
$\displaystyle a A^2 ~\frac{(-A)^{n-2}}{(n-2)!} \left(\ln[z]\right)^{n-2}$-series, $\ldots$,
(see Table 1 in \cite{M97})  into the kernel
which dominate at large $N_f$.

The resummation of this ``chain-1" subseries into an analytic
function in $A$ should not be taken by surprise. Really, the
considered problem can be connected with the calculation of large
$N_f$ asymptotics of ADs' in order of $1/N_f$. An approach was
suggested by A. Vasil'ev and colleagues at the beginning of the
80'es ~\cite{VPH81} to calculate the renormalization-group
functions in this limit, they used the conformal properties of the
theory at the critical point $g=g_c$ corresponding to the
non-trivial zero $g_c$ of the D-dimensional $\beta$-function. This
approach was extended by J. Gracey for the calculation of ADs' of
composite operators of DIS in QCD in any order $n$ of PT,
~\cite{Gr94}. I used another approach which is close to
\cite{P-M-P84}; contrary to the large $N_f$ asymptotic method, it
does not appeal to the value of parameters $N_fT_R$, $C_A/2$ or
$C_F$, associated with different kinds of loops in QCD. To
illustrate this feature, let us consider the insertions of chains
of one-loop self-energy parts into the ``quark" line of diagram
Fig.1a (``chain-2" in \cite{M97}). Contributions of these diagrams
calculated in the framework of the above scalar model
 do not contain the parameter
$N_f$, nevertheless, they can be summarized into the kernel
$P^{(2)}(z; B)$ \cite{M97} \begin{eqnarray}\label{IntB} \displaystyle aP_0(z)= a\bar{z}
\stackrel{chain-2}{\longrightarrow} P^{(2)}(z; B) = a\bar{z} \left(1+B
\frac{d}{dB} \right) \left[ (\bar z)^{-B}
\frac{\gamma_{\psi}(0)}{\gamma_{\psi}(B)} \right];
~\mbox{where}~B=a \gamma_{\psi}(0),~\bar{z}\equiv 1-z, \end{eqnarray}  
according to the same approach. This corresponds to summation of
various series like $(n+1)\cdot
\displaystyle a \frac{(-B)^n}{n!}\left( \ln[\bar{z}]+\frac{5}{3} \right)^n,$ $\ldots$
into the kernel. The operator $\displaystyle \left(1+B~d/dB \right)$
appearing in front of formula (\ref{IntB}) expresses an inherent
combinatoric factor $(n+1)$ to these diagrams. Following that line, the
``improved'' QCD kernel $P^{(1)}(z; A)$ was obtained in
\cite{MS98}  for the general case of a mixed chain (quark and
gluon bubble chain) in $\xi$-- gauge.

Here, we present the QCD results similar to Eq.(\ref{IntA}), in
the covariant $\xi$-- gauge for the DGLAP non-singlet kernel $P(z;
A)$. Analytic properties of the function $P(z; A)$ in variable $A$
are analyzed. The assumption of the ``Naive Nonabelianization''
(NNA) approximation \cite{BrGr95} for the kernel calculation
\cite{GK97} is discussed and its generalization based on $\xi=-3$
gauge is suggested. The numerical importance of the resummation in
this case is demonstrated. The ER-BL evolution kernel $V(x,y)$ is
obtained in the same multiloop approximation as the DGLAP kernel,
by using exact relations between the $P$ and $V$ kernels
\cite{MR85,M97} for a class of ``triangular diagrams''. The
considered class of diagrams represents the leading
$b_0$-contributions to both the kinds of kernels. Partial
solutions for the ER-BL equation, $\Phi_n(x,A)$, are derived. The
multiloop ``improved" kernels $P(z; A)$, $V(x,y; A)$ and solutions
$\Phi_n(x,A)$ are compared with the exact results in 3(2)-loop
approximation.

\section{Triangular diagrams for the DGLAP evolution kernel}
Here, the results of the bubble chain resummation for QCD diagrams
in Fig.1 for the DGLAP kernel are discussed. These classes of
diagrams generate, in particularly, the contributions $\sim a_s
\left(A \ln[1/z] \right)^n/n!$ in any order $n$ of pQCD. Based on the
resummation method of Ref.~\cite{M97} in the QCD version, one can
derive the kernels $P^{(1a,b,c)}$ (corresponding to the diagrams
in Fig.1 a,b,c) in the covariant $\xi-$gauge\footnote{The gauge
parameter $\xi$ is defined via the gluon propagator in the lowest
order $\displaystyle i D_{\mu \nu}(k^2) = \frac{-i
\delta^{ab}}{k^2+i\epsilon} \left(g_{\mu \nu} +
(\xi-1)\frac{k_{\mu}k_{\mu}}{k^2} \right)$}, whose explicit
expressions are presented in ~\cite{MS98}. They contribute to the
total kernel $P^{(1)}(z; A, \xi)$ that has the expected ``plus
form"
\begin{eqnarray}\label{Psum}
 \displaystyle P^{(1)}(z; A, \xi) &=& a_s C_F 2 \cdot
\left[\bar{z} z^{-A}(1-A)^2 + \frac{2 z^{1-A}}{1-z} \right]_{+}
\frac{A(0, \xi)}{A(A, \xi)},\\ \displaystyle a_s P_{0}(z) &=&
a_s C_F 2 \cdot \left[\bar{z} ~~~+ ~~~\frac{2 z}{1-z} \right]_{+}, \end{eqnarray}  

\begin{figure}[th]
\vspace*{-7mm}
\input{psbox.tex}
\noindent
 \hspace*{0.03\textwidth}
  \begin{minipage}{0.9\textwidth}
   $${\psboxto(0.9\textwidth;0cm){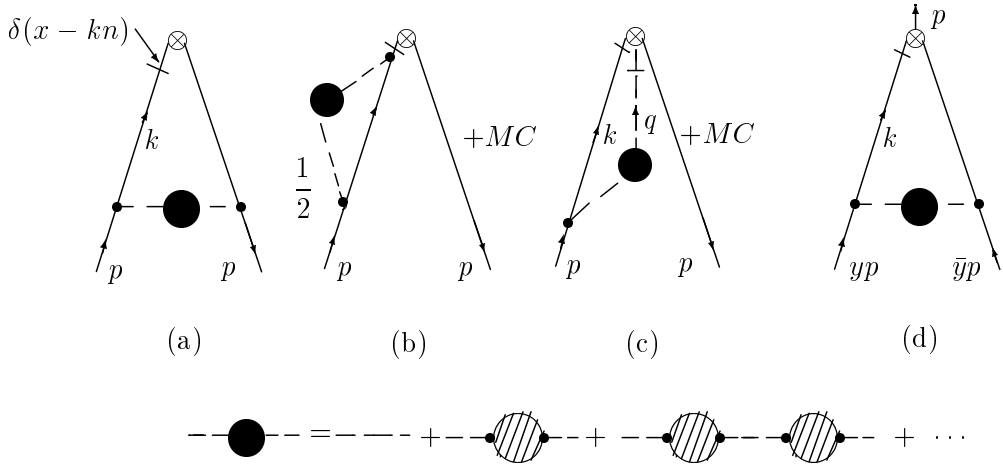}}%
   $$
   \vspace*{-7mm}
     \label{fig:diagr}
     \caption{%
      The diagrams in figs. 1a -- 1c are ``triangular" diagrams
          for the QCD DGLAP kernel; dashed lines for gluons, solid lines for quarks;
          black circles denote the sum of all kinds of the one-loop insertions
         (dashed circles), both quark and gluon (ghost) or mixed chains;
         the slash on the line denotes the delta function $\delta(z-kn)$
         ($k$ is the momentum on the line) which is traced to the representation
         of the composite operator $\otimes$, see~[6] for details;
         MC denotes the mirror--conjugate diagram; 1d is an example of a diagram
         for the nonforward ER-BL kernel.}
 \end{minipage}
\end{figure}
\vspace*{-4mm}

where $\displaystyle a_s=\frac{\alpha_s}{4\pi}$, ~$\displaystyle C_F=
(N_c^2-1)/2N_c$,~$C_A=N_c$, ~$\displaystyle T_R=\frac1{2}$ are the Casimirs
of SU($N_c$) group, quantity $A \equiv A(0,\xi) =-a_s \gamma_g(0,\xi)$, and
the $\gamma_g(0,\xi)$ is the one-loop coefficient of
the standard AD of the gluon field.  For comparison
with (\ref{Psum}), the one-loop result $a_s P_0(z)$ is written
also down, the latter can be obtained as the limit $P^{(1)}(z; A
\to 0, \xi)$. The function $A(\varepsilon,\xi)$ is defined as
$A(\varepsilon,\xi) =-a_s \gamma_g(\varepsilon,\xi)$,
where the
 {\bf function} $\gamma_g(\varepsilon,\xi)$ is the coefficient
of the anomalous dimension in D-dimension, here and below
$D=4-2\varepsilon$. In other words, it is the coefficient
$Z_1(\varepsilon)$ of a simple pole in the expansion of the gluon
field renormalization constant $Z$ that includes both its finite
part and all the powers of the $\varepsilon$-expansion.
 So, one can conclude that the ``all-order" result in
(\ref{Psum}) is completely determined by the expression
$\gamma_g^{(q)}(\varepsilon)$ ($\gamma_g^{(g)}(\varepsilon, \xi)$)
for the single quark (or/and gluon) bubble subgraph.
The function $\gamma_g(\varepsilon,\xi)$ thus
defined is an analytic function in the variable $\varepsilon$ by
construction, see~\cite{M97}. Note that the function
$a_s \gamma_g(\varepsilon,\xi)$ at zero reveals itself again as the
argument of the same function, Eq.(\ref{Psum}), through the variable $A$.
The quantity $A$ here plays the role of a new perturbative expansion parameter.
Equation (\ref{Psum}) is valid for
any kind of insertions, \hbox{\it i.e.}, $\gamma_g = \gamma_g^{(q)}$ for the
quark loop, $\gamma_g = \gamma_g^{(g)}$ for the gluon (ghost)
loop, or for their sum \begin{equation} \label{gamma} \gamma_g(A, \xi) =
\gamma_g^{(q)}(A) + \gamma_g^{(g)}(A, \xi); \end{equation}   when both the
kinds of insertions are taken into account.
The $\delta(1-z)$ - terms appearing in the partial contributions
(see~\cite{MS98}) are exactly accumulated in the form of the
$[\ldots]_{+}$ prescription in (\ref{Psum}),
 and the $\xi$ - terms successfully cancel.
This is due to the evident current conservation for the case of
quark bubble insertions; including the gluon bubbles into
consideration merely modifies the effective AD $\gamma_g^{(q)}(A)
 \to \gamma_g(A,\xi)$, conserving the structure of result
(\ref{Psum}), see ~\cite{M97,MS98}. Substituting  the well-known
expressions of $\gamma_g(\varepsilon)$ from the quark or gluon
(ghost)
 loops (see, \hbox{ \it e.g.}, \cite{IZ})
\begin{eqnarray}\label{Piq}
\gamma_g^{(q)}(\varepsilon) &=& -8 N_f T_R B(D/2,D/2) C(\varepsilon),\\
\gamma_g^{(g)}(\varepsilon, \xi)& =&  \frac{C_A}{2} B(D/2-1,D/2-1) \left(
\left(\frac{3D-2}{D-1}\right) +\right. \nonumber \\
&&  ~~~~~~~~\left.(1-\xi)(D-3) +
 \left(\frac{1-\xi}{2} \right)^2 \varepsilon \right) C(\varepsilon),
\end{eqnarray}into formula (\ref{Psum}) one can obtain $P^{(1)}(z;A,\xi)$
for both the quark and gluon loop insertions simultaneously. Here,
the coefficient
$C(\varepsilon)=\Gamma(1-\varepsilon)\Gamma(1+\varepsilon)$
implies a certain choice of the $\overline{\rm MS}$ scheme where every loop
integral is multiplied by the scheme factor
$\Gamma(D/2-1)(\mu^2/4\pi)^{\varepsilon}$ ($\overline{\rm MS}_1$ scheme).
The renormalization
scheme dependence of $P^{(1)}(z; A)$ is accumulated by the factor
$C(\varepsilon)$. For another popular definition of a
minimal scheme, when a scheme factor is chosen as $\exp(c \cdot
\varepsilon), ~c=-\gamma_E+ \ldots$ instead of
 $\Gamma(D/2-1)$ ($\overline{\rm MS}_2$ scheme), the coefficient $C(\varepsilon)$
 does not contain any scheme ``traces'' in final expressions
for the renormalization-group functions.

Of course, the final result (\ref{Psum}) will be gauge-dependent
in virtue of the evident gauge dependence of the gluon loop
contribution $\gamma_g^{(g)}(\varepsilon, \xi)$. A new expansion
parameter $A$ in this case, \begin{eqnarray}\label{A} A=-a_s
\gamma_g(0, \xi) = -a_s \left( \gamma_g^{(g)}(0, \xi) +
\gamma_g^{(q)}(0)\right)= -a_s
\left[\left(\frac5{3}+\frac{(1-\xi)}{2}\right)C_A
 -  \frac4{3} N_f T_R \right],
\end{eqnarray}is the contribution to the one-loop renormalization of the
gluon field. The positions of zeros of the function $\gamma_g(A,
\xi)$ in $A$, which represent the poles of $P(z; A, \xi)$, also
depend on $\xi$. The kernel  $P^{(1)}(z; A, \xi)$ becomes
gauge-invariant if we restrict themselves only to the quark-loop
insertions, \hbox{\it i.e.}, ~$\gamma_g \to \gamma_g^{(q)}$; ~$\displaystyle A \to
A^{(q)}=-a_s \gamma_g^{(q)}(0) = a_s \frac{4}{3} T_R N_f$, and
$P^{(1)}(z; A, \xi)$ is reduced to $P^{(1)}_q(z; A^{(q)})$, as it
is presented in \cite{M97}. It is instructive to outline analytic
properties of $P^{(1)}_q(z; A^{(q)})$ in $A^{(q)}$ based on Eq.
(\ref{Psum}) and on the explicit form for $\gamma_g^{(q)}$ in
(\ref{Piq}): (i)~the range of convergence of the PT series
corresponds to the left zero
 of the function $\gamma^{(q)}_g(A)$ and is equal to
 $ A_0 = 5/2,$ which corresponds to $\alpha_{s}^0 = 15\pi /N_f$, so,
 this range looks very broad
 \footnote{Here we consider the evolution kernel $P(z, A)$ itself.
We do not consider that the factorization scale $\mu^2$ of hard
processes would be chosen large enough, $\mu^2 \geq m_{\rho}^2$,
where the $\rho$--meson mass $m_{\rho}$ represents the
characteristic hadronic scale. For this reason, the used coupling
$\alpha_s(\mu^2)$ could not be too large.},
  $\alpha_s < 5\pi$ at $N_f=3$;
(ii) the resummation into
 $P^{(1)}_q(z; A)$ is substantial, two zeros of the function
 $P^{(1)}_q(z; A)$ in $A$ appear within the range of convergence
(in $\overline{\rm MS}_1$ scheme). Of course, the moments of this reduced
kernel $P^{(1)}_q(z; A^{(q)})$ agree with the generating function
for the anomalous dimensions, obtained earlier in ~\cite{Gr94}.

\section{A modified NNA version for kernel calculations }
The expansion of $P^{(1)}_q(z; A)$ in $A$ provides the leading
$a_s \cdot \left(a_s N_f \ln[1/z]\right)^n$ dependence of the kernels
with a large number $N_f$ in any order $n$ of PT \cite{M97}. But
these contributions do not numerically dominate for real numbers
of flavours $N_f=4, ~5, ~6$. That can be verified by comparing the
total numerical results for 2-- and 3--loop ADs' of composite
operators (ADCO) presented in \cite{LRV94} with their
$N_f$-leading terms, see  Table 1. There the contributions to
coefficients of different Casimirs in the ADCO are presented. To
obtain a satisfactory agreement at least with the two-loop
results, one should take into account the contribution from {\bf
next-to-leading} $N_f$-terms. As a first step, let us consider the
contribution from the completed renormalization of the gluon line,
which should generate {\bf a part of next-to-leading} terms. Below, we
examine an exceptional choice of the gauge parameter $\xi = -3$.
~For this gauge the coefficient of one-loop gluon AD $\gamma_g(0,
-3)$ coincides with $b_0$, the one-loop coefficient of the
$\beta$-function \footnote{ Here, for the $\beta(a_s)$-function we
adapt $\beta(a_s)= -b_0 a_s^2 - b_1 ~a_s^3 \ldots$, $\displaystyle ~b_0 =
\frac{11}{3}C_A - \frac2{3}N_f,
~b_1=\frac{34}{3}C_A^2 - N_f\left(2C_F+\frac{10}{3}C_A \right),\ldots$} 
and $A=-a_s b_0$.
Therefore this gauge can be used for reformulating the so-called
\cite{BrGr95} NNA proposition to kernel calculations. Note, just
this value of $\xi$ has been used in \cite{Ch96} to estimate the
total gluon contribution only from the gluon bubble in order
$a_s^2$ to the process of $e^+ ~e^-$ annihilation. Other
interesting applications of this gauge to approximate the exact
loop results have been considered in \cite{sirlin95,field98}.

To obtain the NNA result in a usual way, one should substitute the
coefficient $b_0$ for $\gamma^{(q)}_g(0)$ in the expression for
$A^{(q)}$ by hand (see, \hbox{ \it e.g.}, \cite{GK97}). Note, the use of such
an NNA procedure does not improve $P^{(1)}_q(z; A)$ and leads to
poor results even for the two-loop level, \hbox{\it i.e.}, for the
$a_s^2~P_1(z)$ term of the expansion, see ~\cite{MMS97}. The NNA
trick expresses common hope that the main logarithmic contribution
can follow from the renormalization of the coupling constant
$g_s$. The first effective realization of this idea goes back
to the well-known Brodsky-Lepage-Mackenzie (BLM) prescription for the
scale setting \cite{BLM83} formulated in the next-to-leading
order approximation.
That $g_s$-renormalization appears as a sum of contributions from
all the sources of renormalization of $g_s$ at the vertices of
triangular diagrams. Let us consider the gluonic, vertices, and
quark line renormalizations successively in the case of the
$\xi=-3$ gauge. The one-loop gluon renormalization in this gauge
imitates the contributions from all other sources and the
coefficient $b_0$ {\bf appears naturally} via of $\gamma_g(0,-3)$.
At the same time, in the one-loop vertices renormalization constant
$Z_{1F}$,
 $$1-Z_{1F} \sim a_s \left[ C_F \xi + {C_A \over 4}(3+\xi) \right],$$
the nonabelian part vanishes at $\xi=-3$, while the corresponding
Abelian part, $a_s C_F \xi$, is compensated by the renormalization
of the quark line of a triangular diagram, $- a_s C_F \xi$, due to
the abelian Ward identity\footnote{This reason was noted also
in~\cite{field98}}.
So, due to the cancellations, only the gluon
contribution survives in $g_s$ renormalization and provides the
expected $b_0$-term, $a_s b_0\ln[z]$. These properties of
cancellation can be illustrated by the well-known diagram by
diagram results for two-loop $P_{1}(z)$ presented in Feynman gauge
in \cite{FLK81,MR85} (for $V_1(x,y)$ in \cite{MR85,MR86}).
Indeed, the terms, connected with the quark field/vertic
renormalization are proportional to $\ln[1-z]$ in these diagrams
and really cancel in the gauge invariant sum of all contributions.
In contrast to that, the $\ln[z]$-terms collect the coefficient
$a_s b_0$. Though we should not take into account the self-energy
chain (``chain-2" in the Intr.) and ``rainbow" graph insertions
into the quark line unless the vertices of the triangular diagram,
dressed in the same manner, is included into consideration, we see
that their contributions should be cancelled in the first
$log$-parts for the discussed gauge. For these reasons we can
guess the gauge $\xi= -3$  ``exceptional" for the one-loop chain
dressing.

To analyze the resulting effect of ``all-loop" resummation for the case
 $\xi=-3$ in~(\ref{Psum}),
let us choose the common factor $A
/A(A,-3)$ in formula (\ref{P-3}) (below the notation
~$a=a_s b_0 = -A$ is introduced),
\begin{eqnarray} 
\label{P-3}
\displaystyle P^{(1)}(z;-a, -3) = a_s C_F 2 \cdot \left[\bar{z} z^{a}(1+a)^2 + \frac{2
z^{1+a}}{1-z} \right]_{+} \frac{A}{A(-a,
-3)},\\ \displaystyle \frac{A}{A(-a, -3)}\equiv
\frac{b_0}{\gamma_g(-a, -3)} =
\frac{\Gamma(2+2a) (3+2a)}{\left(\Gamma(1+a)\right)^2 C(-a)}
\frac{b_0}{(4C_A a^2 +a (3b_0+2C_A) +3b_0)},
\end{eqnarray}  
 for a crude measure of
the modification of the kernel in comparison with the one-loop
result $a_s P_0(z)$. The factor (as well as the whole kernel $P(z;
-a, -3)$) has no singularity in $a$ for $a > 0$. Considering the
curve of this factor in the argument $a$ in Fig.2, one can
conclude:
\input{psbox}
 \begin{figure}[ht]
  \hspace*{0.0\textwidth}
   \begin{minipage}{1.0 \textwidth}
    $$\psannotate{\psboxto(0.75\textwidth;0cm){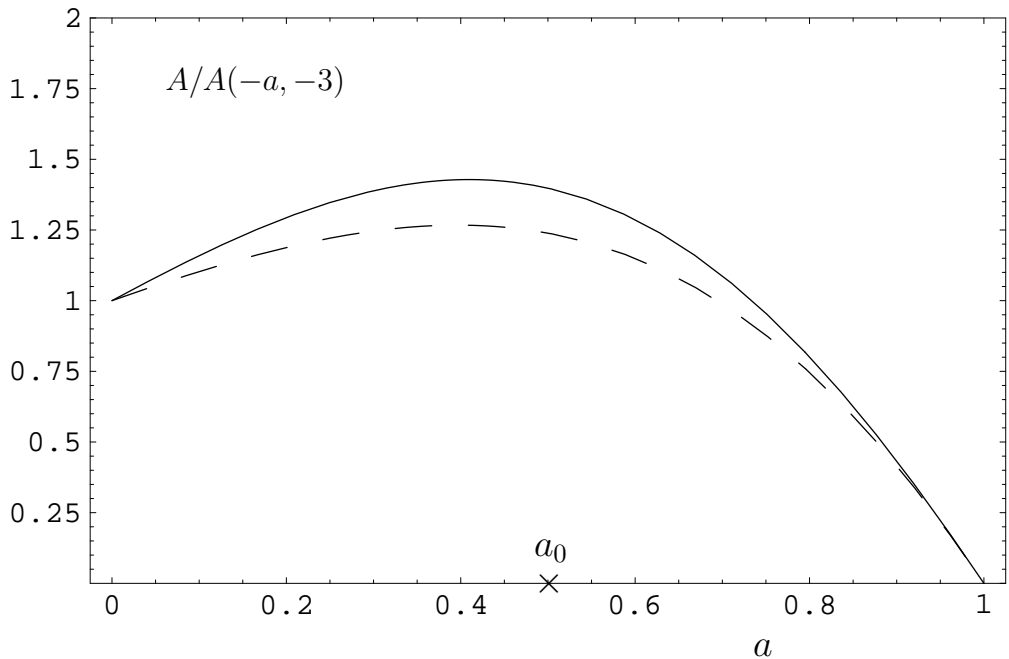}}%
       {\at(11.0\pscm;-0.6\pscm){\Large $a$}}%
        {\at(-6.3\pscm;0.3\pscm){\Large $\times$}}
       {\at(-6.3\pscm;0.8\pscm){\Large $a_0$}}%
       {\at(-11.2\pscm;7\pscm){\large $A/
                                     A(-a,-3)$}}$$
      \label{fig2}
       \caption{The curves of the factor $A/A(-a, -3)$
       in $a$; 
   the solid line corresponds to
   $\overline{\rm MS}_1$ scheme; the dashed line corresponds to $\overline{\rm MS}_2$ scheme}
        \end{minipage}
         \end{figure}
(i) the factor $\displaystyle \frac{A}{A(-a, -3)}$ noticeably
grows with argument $a$ in the range of the standard PT validity.
Really,  this factor reaches
 ~$1.32$ for the $\overline{\rm MS}_1$ scheme ( $1.17$ for the $\overline{\rm MS}_2$ scheme), if we take
the naive boundary of  validity of the standard PT,  $\displaystyle 
a_0=0.5$, ~$\displaystyle \alpha_{s0} =\frac{4 \pi}{b_0}a_0 \approx 0.7$ that
corresponds to the value of $\alpha_s$ on the hadronic scale;
thus, the resummation is numerically important in this range, see
Fig.2.

(ii) scheme dependence looks not too strong for acceptable
     values of parameter $a$.

Note  that Eqs.(\ref{Psum}, \ref{P-3}) could not provide the valid
asymptotic behavior of the kernels for $z \to 0$. A similar
$z$-behavior is determined by the double-logarithmic corrections
which are most singular at zero, like $ a_s \left(a_s \ln^2[z]
\right)^n$~\cite{bv96}. These contributions appear due to
renormalization of the composite operator in the diagrams by
ladder graphs, \hbox{\it etc.}, rather than by the triangular ones. But,
Eq.(\ref{P-3}) can provide a main $z$--behavior for not too small
$z$ due to simple-logarithmic corrections. To obtain the low
boundary of this $z$--region, let us compare effects from  simple
and double logarithmic contributions taking into account the main
singular terms up to 3 loops;
 \begin{eqnarray}\label{Pasym}
 \left|_{z \downarrow 0} P(z)\right. = && a_s 2C_F  \nonumber \\
+&&a_s^2 2C_F \left[ \ldots + b_0
\ln[z]~~~~~~~~~~~~~~~~~~~~~~~~~~~~~~ + (2C_A-3C_F)\ln^2[z] \right]
\nonumber \\ +&&  a_s^3 2 C_F \left[ \ldots +\left( b_0
\frac{11b_0-2}{3}\ln[z]+ \frac{b_0^2}{2} \ln^2[z] \right) +
     \frac{C_F^2 - 6[C_F - C_A/2]^2}{3} \ln^4[z]
     \right] \\
     +&&  \ldots \nonumber  ,
\end{eqnarray}The first terms in the square brackets in (\ref{Pasym}) follow
directly from the expansion\footnote{The expansion of the Eqs.(\ref{P-3},\ref{Psum})
in the next orders generates the Rieman zeta-functions started with
$\zeta(3)$ in order of $O(a_s^4)$}
of Eq.(\ref{P-3}) in $a$; the second
term in the second line is the double-$\log$ from the exact
two-loop calculations; and the last term in the third line was
predicted by J.Blumlein\&A.Vogt in ~\cite{bv96}. From
(\ref{Pasym}) rough estimate follows to the boundary of validity
of Eq.(\ref{P-3}), $z \simeq 0.1 - 0.05$ at moderate $\alpha_s
\simeq 0.3 - 0.1$. The most singular $\ln^4[z]$-term in
(\ref{Pasym}) becomes important for $z \leq 10^{-3}$.
 It seems
naturally to combine the improved by the simple-logs kernel
$P^{(1)}(z;-a,-3)$ with the first double-logs contribution from
the exact two-loop calculations \footnote{Here the double-$log$'s
part is rewritten from \cite{FLK81};
$F(z)=\frac1{2}\ln^2[z]-2\ln[z]\ln[1+z]-2Li_2(-z)-Li_2(1)$ } into
a modernized kernel $\tilde{P}(z)$,$$
\tilde{P}(z)=P^{(1)}(z;-a,-3)+ a_s^2 C_F \left[\left(P_0(z)C_A -
(1+z)C_F \right)\ln^2[z]-4(C_F-C_A/2)P_0(-z)F(z)\right]_+,$$ which
works up to $z \sim 10^{-3}$.

At the end let us consider the integral characteristics of the
kernel $P^{(1)}(z; -a, -3)$ to compare with the exact results.
 The expansion of this kernel in $a$ generates partial
kernels $a_s^2 P_{(1)}(z), \ a_s^3P_{(2)}(z), \ldots$ which in
turn produce ADCO $a_s^2~\Gamma_{(1)}(n)$,
$a_s^3~\Gamma_{(2)}(n)$, $\ldots$ according to the relation
$\Gamma(n) = \int^1_0 dz z^n P(z)$. Let us compare these elements
of ADCO and a few numerical exact results from \cite{LRV94}
collected in Table 1:

(i) evidently, the leading $N_f$-contributions are reproduced
exactly for any $\Gamma_{(j)}(n)$;

(ii) we consider there the next-to-leading $N_f$-contributions to 
the coefficient $\Gamma_{(1)}(n)$ generated by gluon loops and
associated with the Casimirs $C_F C_A/2$, the ~$C_F^2$--term is
missed, but its contribution is  numerically insignificant. It is
seen that in this order the $C_F C_A$--terms are rather close to
exact values (the accuracy is about $10\%$ for $n > 2$) and our
approximation works rather well;

(iii) in the next order, the contributions to $\Gamma_{(2)}(n)$
associated  with the coefficients $N_f \cdot C_F C_A$ and $C_A^2
C_F$ arise, while the terms with the Casimirs  $C_F^3, ~N_f \cdot
C_F^2 ,~C_F^2 C_A$ are missed.
The involved Casimirs ($N_f \cdot C_F C_A$, $C_A^2 C_F$), as it follows
from the estimations in \cite{LRV94}, also dominate numerically in the
third order ADCO $\Gamma_{(2)}(n)$, which gives a hint for success.
Nevertheless, contrary to the
previous item, all the generated terms are opposite in sign to the
exact values, and the ``$\xi=-3$ approximation" doesn't work at
all. So, we need the next step to improve the agreement with
3-loop results -- to obtain the next-to-leading $N_f$-terms by an exact
calculation\footnote {An example of similar calculation in QCD recently has been
demonstrated in \cite{CDGM99}}.
\newpage
 {\bf Table $1$.} The results of $\Gamma_{(1,2)}(n)$ calculations
 ( $\Gamma(n)=\int^1_0 dx x^n P(z)$)
performed in different ways, exact numerical results from \cite{LRV94} and
approximation obtained from $P(z, A, \xi)$ with  $\xi=-3$; both
numerical and analytic {\bf exact results} are marked by the bold
print.

\begin{center}
\begin{tabular}{|c||c|c||c|c|c|}\hline
&\multicolumn{2}{|c||}{{\strut\vphantom{\vbox to 6mm{}}
$\Gamma_{(1)}(n)$ $_{\vphantom{\vbox to 4mm{}}}$}}&
\multicolumn{3}{|c|}{$\Gamma_{(2)}(n)$}\\
\cline{2-6}
 &{\strut\vphantom{\vbox to 6mm{}}
 \hspace{0.1mm}$C_F C_A$\hspace{0.1mm}
   $_{\vphantom{\vbox to 4mm{}}}$}
   &\hspace{1mm}$N_f \cdot C_F$\hspace{1mm}
     &\hspace{1mm}$C_A^2 C_F$\hspace{1mm}
       &\hspace{1mm}$N_f \cdot C_FC_A$\hspace{1mm}
         &\hspace{1mm}$N_f^2 \cdot C_F$\hspace*{1mm}\\
\hline
{\strut\vphantom{\vbox to 6mm{}} n=2 $_{\vphantom{\vbox to
4mm{}}}$}& & & & & \\ {\bf Exact }
 &{\bf 13.9}
   &
     &{\boldmath $\displaystyle 86.1 + 21.3 \ \zeta(3)$}
       &{\boldmath $\displaystyle -12.9-21.3 \ \zeta(3)$}
         &  \\
& &{\boldmath $\displaystyle -2.3704$} & &
&{\boldmath $\displaystyle -0.9218$} \\ 
{\strut\vphantom{\vbox to 6mm{}}
 $\hspace{1mm} \xi=-3 \hspace{1mm}$
  $_{\vphantom{\vbox to 4mm{}}}$}
 &$11.3$
   &
     &$-42.0$
       &$12.9$
         & \\
\hline
{\strut\vphantom{\vbox to 6mm{}}n=4$_{\vphantom{\vbox to
4mm{}}}$:} & & & & & \\ {\bf Exact }
 &{\bf 23.9}
   &
     & {\boldmath $140.0 + 19.2 \ \zeta(3)$}
       &{\boldmath $-18.1 -41.9 \ \zeta(3)$}
         &  \\
& &{\boldmath $\displaystyle -4.9152$} & &
&{\boldmath $\displaystyle -1.5814$} \\ 
{\strut\vphantom{\vbox to 6mm{}} $\hspace{1mm} \xi=-3
\hspace{1mm}$ $_{\vphantom{\vbox to 4mm{}}}$}
 &$23.5$
   &
     &$-76.0$
       &$23.$
         &  \\
\hline
{\strut\vphantom{\vbox to 6mm{}}n=6 $_{\vphantom{\vbox to
4mm{}}}$} & & & & & \\ {\bf Exact }
 &{\bf 29.7}
   &
     &{\boldmath $173+  19.01 \ \zeta(3)$}
       &{\boldmath $-20.4 -54.0 \ \zeta(3) $}
         &  \\
& &{\boldmath $\displaystyle -6.4719$} & &
&{\boldmath $\displaystyle -1.9279$}  \\ 
{\strut\vphantom{\vbox to 6mm{}} $\hspace{1mm} \xi=-3
\hspace{1mm}$ $_{\vphantom{\vbox to 4mm{}}}$}
 &$31.1$
   &
     &$-95.6$
       &$28.5$
         &\\
\hline
{\strut\vphantom{\vbox to 6mm{}}n=8 $_{\vphantom{\vbox to
4mm{}}}$} & & & & & \\ {\bf Exact }
 &{\bf 33.9}
   &
     &{\boldmath $196.9 + 18.98 \ \zeta(3)$}
       &{\boldmath $-21.9 - 62.7 \ \zeta(3)$}
         &  \\
& &{\boldmath $\displaystyle -7.6094$} & &
&{\boldmath $\displaystyle - 2.1619$} \\ 
$\hspace{1mm} \xi=-3 \hspace{1mm}$
 &$36.3$
   &
     &$-109.0$
       &$32.3$
         &  \\
\hline
{\strut\vphantom{\vbox to 6mm{}} n=10 $_{\vphantom{\vbox to
4mm{}}}$}& & & & & \\
 {\bf Exact }
 &{\bf 37.27}
   &
     &{\boldmath $216.0 + 18.96 \ \zeta(3)$}
       &{\boldmath $-23.2 - 69.6 \ \zeta(3)$}
         & \\
& &{\boldmath $-8.5095$} & & &{\boldmath $ -2.3366$} \\ 
$\hspace{1mm} \xi=-3 \hspace{1mm}$
 &$41.00$
   &
     &$-119.28$
       &$35.24$
         &  \\
\hline
{\strut\vphantom{\vbox to 6mm{}} n=12 $_{\vphantom{\vbox to
4mm{}}}$}& & & & & \\
 {\bf Exact }
 &{\bf 40.02}
   &
     &{\bf ?}
       &{\bf ?}
         & \\
& &{\boldmath $-9.2555$} & & &{\boldmath $ -2.4753$} \\ 
$\hspace{1mm} \xi=-3 \hspace{1mm}$
 &$44.64$
   &
     &$-127.61$
       &$37.58$
         &  \\
\hline
\end{tabular}
\end{center}
\vspace{3mm}

On the other hand, it looks rather naive to expect a good agreement
of the values of $\Gamma_{(2)}(n)$, obtained from expansion of (\ref{P-3})
with exact three-loop results.
Really, our approach takes account only of the simplest two-bubble
chain diagrams (to be precise, only 3 types of diagrams) among all the set 
of 3-loop diagrams. What can we expect from the next-to-leading $N_f$ corrections, 
like $a_s \cdot \left(a_s (a_s N_f)^{n} \right)$, mentioned above
in item (iii)? The set of corresponding diagrams starts with {\bf a part} of all
 3-loop diagrams, and this part, as I hope,  dominates at this loop level. 
Let us consider the diagrams underlying the 
``tower" of these corrections; these diagrams contain:

(a) only  one-loop insertion into gluon lines or vertices, the number of such diagrams
in a covariant gauge amounts to 39 (without M.C. diagrams); these
diagrams can be obtained from the set of two-loop diagrams presented,
 \hbox{ \it e.g.}, in \cite{FLK81};

(b) essentially two-loop self-energy insertions into gluon lines, 
they are only of 3 types of the diagrams presented in Fig. 1(a,b,c); now 
the black bubble denotes the sum of these two-loop self-energy parts.

\noindent
The calculation of the contributions from diagrams of type (a) looks as a formidable
task; the result is substantial and could not be guessed {\it a priori}.
On the contrary, diagrams of type (b) lead to partially an expected
contribution to the kernel,
\begin{equation} \label{corr-ad2}
 a_s P_0(z) \cdot a_s^2 \gamma_g^{(1)}(\xi) \ln(z)+ \ldots, 
\end{equation} 
which is evidently generated by the two-loop AD, $a_s^2 \gamma_g^{(1)}(\xi)$,
of the gluon line in Fig. 1, where
\begin{equation} 
\gamma_g^{(1)}(\xi) = \frac{23}{4}C_A^2 -N_f \left(2C_F + \frac{5}{2}C_A \right) - 
\left(\frac{C_A}{2} \right)^2 (\xi -1)\left(\xi + \frac{13}{2}\right),   
\end{equation} 
(see, \hbox{ \it e.g.}, \cite{LV93}). It seems tempting to include that contribution into consideration via
 a modification of basis formula (\ref{Psum}), even though ``by hand".
 Namely, substituting a new ``corrected" expansion parameter $A^*$ for $A$
   $$A \to A^*= -a_s \gamma_g(0, \xi) - a_s^2  \gamma_g^{(1)}(\xi) $$
    into expression (\ref{Psum}), one can
     restore the contribution (\ref{corr-ad2}) in the expansion of this
     model $P^{(1)}(z; A^*, \xi)$ kernel.
As the next step, one should choose a new 
 ``corrected" value of the gauge parameter, $\xi^* = -3 + O(a_s)\to \xi$.
  Following the NNA idea and our previous reasoning about an exceptional $\xi=-3$ gauge, 
let us define it by a natural condition  through the $\beta$-function
\begin{eqnarray} 
\gamma_g(0, \xi^*) + a_s \gamma_g^{(1)}(\xi^*) = b_0 + a_s b_1 + O(a_s^2),
\end{eqnarray}  
that leads to the value 
$$\xi^* = -3 + a_s \frac{5}{3} C_A\left(N_f - \frac{5}{2}C_A \right) + O(a_s^2).$$
The hypothesis on $P^{(1)}(z; A^*, \xi^*)$ only slightly reduces the 
discrepancy between the exact and
model results for $\Gamma_{(2)}(n)$ in Table 1. Moreover, it generates 
 the contributions to a new required Casimir $N_f \cdot C_F^2$, which appear of the same 
sign, and are in order smaller
than the exact ones. It is clear, that the model $P^{(1)}(z; A^*, \xi^*)$ is a step along
the right direction, but it is obviously insufficient. 
So, we insist on accurate calculations of  both the types (a) and (b)
diagrams to obtain a reasonable approximation to exact $\Gamma_{(2)}(n)$-results.

\section{The nonforward ER-BL evolution equation and its \\solution}
Here we present the results of the bubble resummation for the
ER-BL kernel $V(x,y)$. The latter can be derived in the same
manner as it was done for the DGLAP kernel $P(z)$, see Appendix A
in \cite{M97}. On the other hand, $V(x,y)$ can be obtained as a
``by-product" of the previous results for  $P(z)$, \hbox{\it i.e.},  we use
again \cite{M97,MS98} the exact relations between the $V$ and $P$
kernels established in any order of PT \cite{MR85} for triangular
diagrams. These relations were obtained by comparing counterparts
for the same triangular diagrams considered in ``forward", Fig.1a,
and ``nonforward",  Fig.1d, kinematics.

Collecting the contributions from triangular diagrams, see
~\cite{MS98}, one arrives at the final expression for $V^{(1)}$ in
the ``main bubbles'' approximation \begin{eqnarray}\label{Vsum} V^{(1)}(x,y; A,
\xi) = a_s C_F 2 \left[ \theta(y > x) \left( \frac{x}{y}
\right)^{1-A} \left(1-A + \frac{1}{y-x} \right)
 \right]_{+} \frac{A(0, \xi)}{A(A, \xi)}+(x\to
 \bar{x},y\to \bar{y})
\end{eqnarray}that has a ``plus form'' again due to the vector current
conservation. The contribution $V^{(1)}$ in (\ref{Vsum}) should
dominate for $N_f \gg 1$ in the kernel $V$. Besides, the function
$V^{(1)}(x,y;A, \xi)$ possesses an important symmetry of its
arguments $x$ and $y$. Indeed, the function ${\cal V}(x,y; A,
\xi)=V^{(1)}(x,y; A, \xi) \cdot (\bar{y} y)^{1-A}$ is symmetric under
the change $x \leftrightarrow y$, ${\cal V}(x,y)={\cal V}(y,x)$.
This symmetry allows us to obtain the eigenfunctions $\psi_n(x)$
of the ``reduced'' evolution equation \cite{MR86} \begin{eqnarray}\label{ev}
&&\int\limits_{0}^{1} V^{(1)}(x,y; A)\psi_n(y;A) dy= \Gamma(n;A)
\psi_n(x;A), \\ \label{solv0} &&\displaystyle \psi_n(y;A) = (\bar{y} y)^{d_{\psi}(A)
- \frac1{2}} ~\frac{C_{n}^{\left(d_{\psi}(A)\right)}
(y-\bar{y})}{N(n,A)}, ~\mbox{here} ~~d_{\psi}(A) = (D_A-1)/2,
~~D_A=4-2A, \\ && N(n,A)=2^{1-4d_{\psi}(A)} \pi
\Gamma(n+2d_{\psi}(A))/
 \left( n!~(n+d_{\psi}(A)) \left(\Gamma(d_{\psi}(A))^2 \right)
 \right), \nonumber
 \label{solv1} \end{eqnarray}  
 and $d_{\psi}(A)$ is the
effective dimension of the quark field when the AD $A$ is
taken into account; $C_{n}^{(\alpha)}(z)$ are the Gegenbauer
polynomials of an order of $\alpha$; $N(n,A)$ is the norm of
$C_{n}^{(\alpha)}(y-\bar{y})$, \cite{BE}. The partial solutions
$\Phi(x; a_s, l)$ of the original ER-BL--equation ( where $l
\equiv \ln(\mu^2/\mu_0^2)$) \begin{equation} \label{BL} \mu^2 \frac{d}{d \mu^2}
 \Phi(x; a_s, l)
= \int^1_0 V^{(1)}(x,y; A)~\Phi(y; a_s, l) dy \end{equation}  are proportional
to these eigenfunctions $\psi_n(x; A)$ for a special case of the
stopped evolution $a_s=a_s^{\star}$, $~\beta(a_s^{\star})=0$, see,
\hbox{ \it e.g.}, ~\cite{Mu95,M97}.
The result (\ref{solv0}) for the eigenfunctions at $\xi=-3$,
 has been confirmed in ~\cite{BeMul97} by
``a partial resummation of conformal anomalies" and in a
suggestion of a large value of $b_0$. Let us examine
$\psi_n(x;-a)$ in (\ref{solv0}) as an approximation to the exact
two-loop solution
 derived in a closed form in \cite{Mu95}.
Expanding, \hbox{ \it e.g.}, $\psi_0(y;-a)$ in parameter $a$ we can express
$\psi_0^{appr}(x)$ versus the exact solution $\psi_0^{exact}(x)$
\begin{eqnarray}\label{approx}
 \psi_0(x;-a) \to
 \psi_0^{appr}(x)&=&6x\bar{x}
\left\{1+ a_s
      b_0 \left(\ln(x \bar{x})+ \frac{5}{3}\right) \right\},\\
 \label{exact}
 \psi_0^{exact}(x)&=&6x \bar{x} \left\{1+ a_s
 b_0\left(\ln(x \bar{x})+ \frac{5}{3}\right) + a_s C_F\left(
 \ln^2\left(\frac{\bar{x}}{x}\right)+ 2- \frac{\pi^2}{2}\right)
 \right\}.
\end{eqnarray}The term  $ \psi_0^{appr}(x)$ coincides
 with the ``conformal symmetry-predicted" (CSP)
 part in (\ref{exact}),
( proportional to $b_0$), this part dominates in
$\psi_0^{exact}(x)$ in the mid-region of the parameter $x$,
$0.3<x<0.7$. The other part in (\ref{exact}) is generated by the
``additional conformal symmetry breaking term" \cite{Mu95}; it
contributes in the opposite phase to the first one and it is large
and enhanced near the end points. For the latter reason, $
\psi_n^{appr}(x)$ become useless at $n \geq 2$ even for the
mid-region $x$ description, see ~\cite{Mu95}.

 In the general case $\beta (a_s) \not=0$ let us start with
an ansatz for the partial solution of Eq.(\ref{BL}), $\Phi_n(x;
a_s, l)$ $\sim \chi_n(a_s, l) \cdot \psi_n(x; A)$, with the
boundary condition $\chi_n(a_s, 0)=1$; $\Phi_n(x; a_s, 0) \sim
\psi_n(x; A)$. For this ansatz, Eq.(\ref{BL}) reduces to
 \begin{equation} \label{BLn}
\left(\mu^2 \partial_{\mu^2} + \beta(a_s)\partial_{a_s}\right)
\ln\left(\Phi_n(x; a_s, l)\right) = \Gamma(n; A). \end{equation}  In the case
$n=0$, the AD of the vector current ~$\Gamma(0; A)=0$, and the
solution of the homogeneous equation in (\ref{BLn}) provides the
``asymptotic wave function"
\begin{equation} 
\label{asympsol} \displaystyle \Phi_0(x; a_s, l) = \psi_0(x; \bar{A}) =
\frac{1}{N(0,\bar{A})}((1-x) x)^{(1-\bar{A})}
 \end{equation} 
 where
$\bar{A} = - \bar{a}_s(\mu^2) \gamma(0,\xi)$, and $\bar{a}_s(\mu^2)$
is the running coupling corresponding to a $\beta$-function
$\beta(a_s)$. Similar solutions have been discussed in
~\cite{GK97} in the framework of the standard NNA approximation.
 Solving simultaneously Eq. (\ref{BLn}) and the
renormalization-group equation for the coupling constant
$\bar{a}_s$, we arrive at the partial solution
$\Phi_n(x;\bar{a}_s, l)$ in the form \begin{equation} \label{BLsol}
 \Phi_n(x,\bar{a}_s) \sim
\chi_n(\mu^2) \cdot \psi_n(x;\bar A); ~\mbox{where} ~\chi_n(\mu^2)
= \exp \left\{- \int^{a_s(\mu^2)}_{a_s(\mu_0^2)}
\frac{\Gamma(n,A)}{\beta(a)}da\right\}. \end{equation}  An adequate choice of
$\beta$-function in (\ref{BLsol}) must correspond to the same
modified NNA approximation that was applied for $\Gamma(n,A)$
calculation, but it is absent yet. The $\beta$-function in a large
$N_f$ expansion, that is equivalent to quark bubbles resummation,
has been computed in ~\cite{Gr96}.

\section{Conclusion}
Here, I present closed expressions in the ``all order"
approximation for the DGLAP kernel $P(z)$ and ER-BL kernel
$V(x,y)$ resulting from resummation of a certain class of QCD
diagrams with the renormalon chain insertions. The contributions
from these diagrams, $P^{(1)}(z; A)$ and $V^{(1)}(z; A)$, give the
leading $N_f$ dependence of the kernels for a large number of
flavours $N_f \gg 1$. These multiloop ``improved" kernels are
generating functions to obtain contributions to partial kernels
like $a_s^{(n+1)}P_{(n)}(z)$ in any order $n$ of the perturbation
expansion. Here $A \sim a_s$ is the new expansion parameter that
coincides (in magnitude) with the anomalous dimension of the gluon
field. On the other hand, the method of calculation suggested in
~\cite{M97} does not depend on the nature of self-energy
insertions and does not appeal to the value of parameters $N_f
T_R, ~C_A/2$ or ~$C_F$ associated with different loops. This
allows us to obtain contributions from chains with different kinds
of self-energy insertions, both quark and gluon (ghost) loops, see
\cite{MS98}. The price for this generalization is the gauge
dependence of final results for $P^{(1)}(z;A(\xi),\xi)$
 and $V^{(1)}(z; A(\xi),\xi)$ on the gauge parameter $\xi$.

The result for the DGLAP nonsinglet kernel  $P^{(1)}(z; A,
\xi)$ is presented in (\ref{Psum}) in the covariant $\xi$-gauge,
it looks similar in form to the simple one-loop kernel. The
analytic properties of this kernel in the variable $a_s$  are
discussed for an exceptional gauge parameter $\xi= -3$. This
choice of the gauge allows one to generalize the naive
nonabelianization suggestion and provides the leading
$b_0$-behavior of the kernel for large $b_0 \gg 1$. For this gauge
$P^{(1)}(z; A, -3)$ in (\ref{P-3}) works up to $z \simeq
0.1-0.05$ at moderate $\alpha_s = 0.3-0.1$, and reproduces
two-loop anomalous dimensions $ a_s^2 \Gamma_{(1)}(n)$ with a good
accuracy, while the standard ``naive nonabelianization"
proposition fails at this level.  But on the next three loop level
the ``$ \xi=-3$ approximation" is insufficient, see quantities
$\Gamma_{(2)}(n)$ in Table 1. At the end, a hypothesis about a possibility
to extend the approach to 3-loop level is briefly discussed.

The contribution $V^{(1)}(x,y; A,\xi)$ to the nonforward
ER-BL kernel (\ref{Vsum}) is obtained for the same classes of
diagrams as a ``byproduct" of the previous technique
\cite{MS98,MR85}. The partial solutions (\ref{ev}), (\ref{BLsol})
to the multiloop improved ER-BL equation are derived, that are
similar in form to the one-loop solutions. The form of these
solutions appearing at  $\xi =-3$ was confirmed independently in
\cite{BeMul97}. The lowest harmonic $\psi_{0}(x;\bar{A})$
roughly imitates the $x$-behavior in the mid-region of the exact
two-loop solution ( \cite{Mu95}).

The obtained results are certainly useful for an independent check
of complicated computer calculations in higher orders of
perturbation theory, similar to \cite{LRV94}; they are useful for
the analysis of evolution ``at small $x$"; they may be a starting
point for further multiloop approximation procedures. \vspace{2mm}

{\bf Acknowledgements} The author is grateful to Dr. K. Chetyrkin,
Dr. A. Grozin, Dr. D. M\"uller, Dr.Dr. N. Stefanis for fruitful
discussions of the results and Dr. A. Kotikov and the referee of 
this paper for careful reading
of manuscript and useful remarks . This investigation has been
supported by the Russian Foundation for Basic Research (RFBR)
98-02-16923.

\end{document}

%% file: psbox.tex
\def\temp{1.34}%
\let\tempp=\relax
\expandafter\ifx\csname psboxversion\endcsname\relax
  \message{PSBOX(\temp) loading}%
\else
    \ifdim\temp cm>\psboxversion cm
      \message{PSBOX(\temp) loading}%
    \else
      \message{PSBOX(\psboxversion) is already loaded: I won't load
        PSBOX(\temp)!}%
      \let\temp=\psboxversion
      \let\tempp= 
    \fi
\fi
\tempp
\let\psboxversion=\temp
\catcode`\@=11
%
%
\def\psfortextures{
\def\PSspeci@l##1##2{%
\special{illustration ##1\space scaled ##2}%
}}%
\def\psfordvitops{
\def\PSspeci@l##1##2{%
\special{dvitops: import ##1\space \the\drawingwd \the\drawinght}%
}}%
\def\psfordvips{
\def\PSspeci@l##1##2{%
\d@my=0.1bp \d@mx=\drawingwd \divide\d@mx by\d@my
\includegraphics{##1\space}}}%
\def\psforoztex{
\def\PSspeci@l##1##2{%
\special{##1 \space
      ##2 1000 div dup scale
      \number-\psllx\space \number-\pslly\space translate
}}}%
\def\psfordvitps{
\def\psdimt@n@sp##1{\d@mx=##1\relax\edef\psn@sp{\number\d@mx}}
\def\PSspeci@l##1##2{%
\special{dvitps: Include0 "psfig.psr"}
\psdimt@n@sp{\drawingwd}
\special{dvitps: Literal "\psn@sp\space"}
\psdimt@n@sp{\drawinght}
\special{dvitps: Literal "\psn@sp\space"}
\psdimt@n@sp{\psllx bp}
\special{dvitps: Literal "\psn@sp\space"}
\psdimt@n@sp{\pslly bp}
\special{dvitps: Literal "\psn@sp\space"}
\psdimt@n@sp{\psurx bp}
\special{dvitps: Literal "\psn@sp\space"}
\psdimt@n@sp{\psury bp}
\special{dvitps: Literal "\psn@sp\space startTexFig\space"}
\special{dvitps: Include1 "##1"}
\special{dvitps: Literal "endTexFig\space"}
}}%
\def\psfordvialw{
\def\PSspeci@l##1##2{
\special{language "PostScript",
position = "bottom left",
literal "  \psllx\space \pslly\space translate
  ##2 1000 div dup scale
  -\psllx\space -\pslly\space translate",
include "##1"}
}}%
\def\psforptips{
\def\PSspeci@l##1##2{{
\d@mx=\psurx bp
\advance \d@mx by -\psllx bp
\divide \d@mx by 1000\multiply\d@mx by \xscale
\incm{\d@mx}
\let\tmpx\dimincm
\d@my=\psury bp
\advance \d@my by -\pslly bp
\divide \d@my by 1000\multiply\d@my by \xscale
\incm{\d@my}
\let\tmpy\dimincm
\d@mx=-\psllx bp
\divide \d@mx by 1000\multiply\d@mx by \xscale
\d@my=-\pslly bp
\divide \d@my by 1000\multiply\d@my by \xscale
\at(\d@mx;\d@my){\special{ps:##1 x=\tmpx, y=\tmpy}}
}}}%
\def\psonlyboxes{
\def\PSspeci@l##1##2{%
\at(0cm;0cm){\boxit{\vbox to\drawinght
  {\vss\hbox to\drawingwd{\at(0cm;0cm){\hbox{({\tt##1})}}\hss}}}}
}}%
\def\psloc@lerr#1{%
\let\savedPSspeci@l=\PSspeci@l%
\def\PSspeci@l##1##2{%
\at(0cm;0cm){\boxit{\vbox to\drawinght
  {\vss\hbox to\drawingwd{\at(0cm;0cm){\hbox{({\tt##1}) #1}}\hss}}}}
\let\PSspeci@l=\savedPSspeci@l
}}%
%
%
\newread\pst@mpin
\newdimen\drawinght\newdimen\drawingwd
\newdimen\psxoffset\newdimen\psyoffset
\newbox\drawingBox
\newcount\xscale \newcount\yscale \newdimen\pscm\pscm=1cm
\newdimen\d@mx \newdimen\d@my
\newdimen\pswdincr \newdimen\pshtincr
\let\ps@nnotation=\relax
{\catcode`\|=0 |catcode`|\=12 |catcode`|
|catcode`#=12 |catcode`*=14
|xdef|backslashother{\}*
|xdef|percentother{
|xdef|tildeother{~}*
|xdef|sharpother{#}*
}%
\def\R@moveMeaningHeader#1:->{}%
\def\uncatcode#1{%
\edef#1{\expandafter\R@moveMeaningHeader\meaning#1}}%
\def\execute#1{#1}
\def\psm@keother#1{\catcode`#112\relax}
\def\executeinspecs#1{%
\execute{\begingroup\let\do\psm@keother\dospecials\catcode`\^^M=9#1\endgroup}}%
\def\@mpty{}%
\def\matchexpin#1#2{
  \fi%
  \edef\tmpb{{#2}}%
  \expandafter\makem@tchtmp\tmpb%
  \edef\tmpa{#1}\edef\tmpb{#2}%
  \expandafter\expandafter\expandafter\m@tchtmp\expandafter\tmpa\tmpb\endm@tch%
  \if\match%
}%
\def\matchin#1#2{%
  \fi%
  \makem@tchtmp{#2}%
  \m@tchtmp#1#2\endm@tch%
  \if\match%
}%
\def\makem@tchtmp#1{\def\m@tchtmp##1#1##2\endm@tch{%
  \def\tmpa{##1}\def\tmpb{##2}\let\m@tchtmp=\relax%
  \ifx\tmpb\@mpty\def\match{YN}%
  \else\def\match{YY}\fi%
}}%
\def\incm#1{{\psxoffset=1cm\d@my=#1
 \d@mx=\d@my
  \divide\d@mx by \psxoffset
  \xdef\dimincm{\number\d@mx.}
  \advance\d@my by -\number\d@mx cm
  \multiply\d@my by 100
 \d@mx=\d@my
  \divide\d@mx by \psxoffset
  \edef\dimincm{\dimincm\number\d@mx}
  \advance\d@my by -\number\d@mx cm
  \multiply\d@my by 100
 \d@mx=\d@my
  \divide\d@mx by \psxoffset
  \xdef\dimincm{\dimincm\number\d@mx}
}}%
%
\newif\ifNotB@undingBox
\newhelp\PShelp{Proceed: you'll have a 5cm square blank box instead of
your graphics (Jean Orloff).}%
\def\s@tsize#1 #2 #3 #4\@ndsize{
  \def\psllx{#1}\def\pslly{#2}%
  \def\psurx{#3}\def\psury{#4}
  \ifx\psurx\@mpty\NotB@undingBoxtrue
  \else
    \drawinght=#4bp\advance\drawinght by-#2bp
    \drawingwd=#3bp\advance\drawingwd by-#1bp
  \fi
  }%
\def\sc@nBBline#1:#2\@ndBBline{\edef\p@rameter{#1}\edef\v@lue{#2}}%
\def\g@bblefirstblank#1#2:{\ifx#1 \else#1\fi#2}%
{\catcode`\%=12
\xdef\B@undingBox{
\def\ReadPSize#1{
 \readfilename#1\relax
 \let\PSfilename=\lastreadfilename
 \openin\pst@mpin=#1\relax
 \ifeof\pst@mpin \errhelp=\PShelp
   \errmessage{I haven't found your postscript file (\PSfilename)}%
   \psloc@lerr{was not found}%
   \s@tsize 0 0 142 142\@ndsize
   \closein\pst@mpin
 \else
   \if\matchexpin{\GlobalInputList}{, \lastreadfilename}%
   \else\xdef\GlobalInputList{\GlobalInputList, \lastreadfilename}%
     \immediate\write\psbj@inaux{\lastreadfilename,}%
   \fi%
   \loop
     \executeinspecs{\catcode`\ =10\global\read\pst@mpin to\n@xtline}%
     \ifeof\pst@mpin
       \errhelp=\PShelp
       \errmessage{(\PSfilename) is not an Encapsulated PostScript File:
           I could not find any \B@undingBox: line.}%
       \edef\v@lue{0 0 142 142:}%
       \psloc@lerr{is not an EPSFile}%
       \NotB@undingBoxfalse
     \else
       \expandafter\sc@nBBline\n@xtline:\@ndBBline
       \ifx\p@rameter\B@undingBox\NotB@undingBoxfalse
         \edef\t@mp{%
           \expandafter\g@bblefirstblank\v@lue\space\space\space}%
         \expandafter\s@tsize\t@mp\@ndsize
       \else\NotB@undingBoxtrue
       \fi
     \fi
   \ifNotB@undingBox\repeat
   \closein\pst@mpin
 \fi
\message{#1}%
}%
%
%
\def\psboxto(#1;#2)#3{\vbox{%
   \ReadPSize{#3}%
   \advance\pswdincr by \drawingwd
   \advance\pshtincr by \drawinght
   \divide\pswdincr by 1000
   \divide\pshtincr by 1000
   \d@mx=#1
   \ifdim\d@mx=0pt\xscale=1000
         \else \xscale=\d@mx \divide \xscale by \pswdincr\fi
   \d@my=#2
   \ifdim\d@my=0pt\yscale=1000
         \else \yscale=\d@my \divide \yscale by \pshtincr\fi
   \ifnum\yscale=1000
         \else\ifnum\xscale=1000\xscale=\yscale
                    \else\ifnum\yscale<\xscale\xscale=\yscale\fi
              \fi
   \fi
   \divide\drawingwd by1000 \multiply\drawingwd by\xscale
   \divide\drawinght by1000 \multiply\drawinght by\xscale
   \divide\psxoffset by1000 \multiply\psxoffset by\xscale
   \divide\psyoffset by1000 \multiply\psyoffset by\xscale
   \global\divide\pscm by 1000
   \global\multiply\pscm by\xscale
   \multiply\pswdincr by\xscale \multiply\pshtincr by\xscale
   \ifdim\d@mx=0pt\d@mx=\pswdincr\fi
   \ifdim\d@my=0pt\d@my=\pshtincr\fi
   \message{scaled \the\xscale}%
 \hbox to\d@mx{\hss\vbox to\d@my{\vss
   \global\setbox\drawingBox=\hbox to 0pt{\kern\psxoffset\vbox to 0pt{%
      \kern-\psyoffset
      \PSspeci@l{\PSfilename}{\the\xscale}%
      \vss}\hss\ps@nnotation}%
   \global\wd\drawingBox=\the\pswdincr
   \global\ht\drawingBox=\the\pshtincr
   \global\drawingwd=\pswdincr
   \global\drawinght=\pshtincr
   \baselineskip=0pt
   \copy\drawingBox
 \vss}\hss}%
  \global\psxoffset=0pt
  \global\psyoffset=0pt
  \global\pswdincr=0pt
  \global\pshtincr=0pt 
  \global\pscm=1cm 
}}%
%
%
\def\psboxscaled#1#2{\vbox{%
  \ReadPSize{#2}%
  \xscale=#1
  \message{scaled \the\xscale}%
  \divide\pswdincr by 1000 \multiply\pswdincr by \xscale
  \divide\pshtincr by 1000 \multiply\pshtincr by \xscale
  \divide\psxoffset by1000 \multiply\psxoffset by\xscale
  \divide\psyoffset by1000 \multiply\psyoffset by\xscale
  \divide\drawingwd by1000 \multiply\drawingwd by\xscale
  \divide\drawinght by1000 \multiply\drawinght by\xscale
  \global\divide\pscm by 1000
  \global\multiply\pscm by\xscale
  \global\setbox\drawingBox=\hbox to 0pt{\kern\psxoffset\vbox to 0pt{%
     \kern-\psyoffset
     \PSspeci@l{\PSfilename}{\the\xscale}%
     \vss}\hss\ps@nnotation}%
  \advance\pswdincr by \drawingwd
  \advance\pshtincr by \drawinght
  \global\wd\drawingBox=\the\pswdincr
  \global\ht\drawingBox=\the\pshtincr
  \global\drawingwd=\pswdincr
  \global\drawinght=\pshtincr
  \baselineskip=0pt
  \copy\drawingBox
  \global\psxoffset=0pt
  \global\psyoffset=0pt
  \global\pswdincr=0pt
  \global\pshtincr=0pt 
  \global\pscm=1cm
}}%
%
\def\psbox#1{\psboxscaled{1000}{#1}}%
\newif\ifn@teof\n@teoftrue
\newif\ifc@ntrolline
\newif\ifmatch
\newread\j@insplitin
\newwrite\j@insplitout
\newwrite\psbj@inaux
\immediate\openout\psbj@inaux=psbjoin.aux
\immediate\write\psbj@inaux{\string\joinfiles}%
\immediate\write\psbj@inaux{\jobname,}%
%
%
\def\toother#1{\ifcat\relax#1\else\expandafter%
  \toother@ux\meaning#1\endtoother@ux\fi}%
\def\toother@ux#1 #2#3\endtoother@ux{\def\tmp{#3}%
  \ifx\tmp\@mpty\def\tmp{#2}\let\next=\relax%
  \else\def\next{\toother@ux#2#3\endtoother@ux}\fi%
\next}%
%
%
\let\readfilenamehook=\relax
\def\re@d{\expandafter\re@daux}
\def\re@daux{\futurelet\nextchar\stopre@dtest}%
\def\re@dnext{\xdef\lastreadfilename{\lastreadfilename\nextchar}%
  \afterassignment\re@d\let\nextchar}%
\def\stopre@d{\egroup\readfilenamehook}%
\def\stopre@dtest{%
  \ifcat\nextchar\relax\let\nextread\stopre@d
  \else
    \ifcat\nextchar\space\def\nextread{%
      \afterassignment\stopre@d\chardef\nextchar=`}%
    \else\let\nextread=\re@dnext
      \toother\nextchar
      \edef\nextchar{\tmp}%
    \fi
  \fi\nextread}%
\def\readfilename{\bgroup%
  \let\\=\backslashother \let\%=\percentother \let\~=\tildeother
  \let\#=\sharpother \xdef\lastreadfilename{}%
  \re@d}%
%
%
\xdef\GlobalInputList{\jobname}%
\def\psnewinput{%
  \def\readfilenamehook{
    \if\matchexpin{\GlobalInputList}{, \lastreadfilename}%
    \else\xdef\GlobalInputList{\GlobalInputList, \lastreadfilename}%
      \immediate\write\psbj@inaux{\lastreadfilename,}%
    \fi%
    \ps@ldinput\lastreadfilename\relax%
    \let\readfilenamehook=\relax%
  }\readfilename%
}%
\expandafter\ifx\csname @@input\endcsname\relax    
  \immediate\let\ps@ldinput=\input\def\input{\psnewinput}%
\else
  \immediate\let\ps@ldinput=\@@input
  \def\@@input{\psnewinput}%
\fi%
\def\nowarnopenout{%
 \def\warnopenout##1##2{%
   \readfilename##2\relax
   \message{\lastreadfilename}%
   \immediate\openout##1=\lastreadfilename\relax}}%
\def\warnopenout#1#2{%
 \readfilename#2\relax
 \def\t@mp{TrashMe,psbjoin.aux,psbjoint.tex,}\uncatcode\t@mp
 \if\matchexpin{\t@mp}{\lastreadfilename,}%
 \else
   \immediate\openin\pst@mpin=\lastreadfilename\relax
   \ifeof\pst@mpin
     \else
     \errhelp{If the content of this file is so precious to you, abort (ie
press x or e) and rename it before retrying.}%
     \errmessage{I'm just about to replace your file named \lastreadfilename}%
   \fi
   \immediate\closein\pst@mpin
 \fi
 \message{\lastreadfilename}%
 \immediate\openout#1=\lastreadfilename\relax}%
{\catcode`\%=12\catcode`\*=14
\gdef\splitfile#1{*
 \readfilename#1\relax
 \immediate\openin\j@insplitin=\lastreadfilename\relax
 \ifeof\j@insplitin
   \message{! I couldn't find and split \lastreadfilename!}*
 \else
   \immediate\openout\j@insplitout=TrashMe
   \message{< Splitting \lastreadfilename\space into}*
   \loop
     \ifeof\j@insplitin
       \immediate\closein\j@insplitin\n@teoffalse
     \else
       \n@teoftrue
       \executeinspecs{\global\read\j@insplitin to\spl@tinline\expandafter
         \ch@ckbeginnewfile\spl@tinline
       \ifc@ntrolline
       \else
         \toks0=\expandafter{\spl@tinline}*
         \immediate\write\j@insplitout{\the\toks0}*
       \fi
     \fi
   \ifn@teof\repeat
   \immediate\closeout\j@insplitout
 \fi\message{>}*
}*
\gdef\ch@ckbeginnewfile#1
 \def\t@mp{#1}*
 \ifx\@mpty\t@mp
   \def\t@mp{#3}*
   \ifx\@mpty\t@mp
     \global\c@ntrollinefalse
   \else
     \immediate\closeout\j@insplitout
     \warnopenout\j@insplitout{#2}*
     \global\c@ntrollinetrue
   \fi
 \else
   \global\c@ntrollinefalse
 \fi}*
\gdef\joinfiles#1\into#2{*
 \message{< Joining following files into}*
 \warnopenout\j@insplitout{#2}*
 \message{:}*
 {*
 \edef\w@##1{\immediate\write\j@insplitout{##1}}*
\w@{
\w@{
\w@{
\w@{
\w@{
\w@{
\w@{
\w@{
\w@{
\w@{
\w@{\string\input\space psbox.tex}*
\w@{\string\splitfile{\string\jobname}}*
\w@{\string\let\string\autojoin=\string\relax}*
}*
 \expandafter\tre@tfilelist#1, \endtre@t
 \immediate\closeout\j@insplitout
 \message{>}*
}*
\gdef\tre@tfilelist#1, #2\endtre@t{*
 \readfilename#1\relax
 \ifx\@mpty\lastreadfilename
 \else
   \immediate\openin\j@insplitin=\lastreadfilename\relax
   \ifeof\j@insplitin
     \errmessage{I couldn't find file \lastreadfilename}*
   \else
     \message{\lastreadfilename}*
     \immediate\write\j@insplitout{
     \executeinspecs{\global\read\j@insplitin to\oldj@ininline}*
     \loop
       \ifeof\j@insplitin\immediate\closein\j@insplitin\n@teoffalse
       \else\n@teoftrue
         \executeinspecs{\global\read\j@insplitin to\j@ininline}*
         \toks0=\expandafter{\oldj@ininline}*
         \let\oldj@ininline=\j@ininline
         \immediate\write\j@insplitout{\the\toks0}*
       \fi
     \ifn@teof
     \repeat
   \immediate\closein\j@insplitin
   \fi
   \tre@tfilelist#2, \endtre@t
 \fi}*
}%
\def\autojoin{%
 \immediate\write\psbj@inaux{\string\into{psbjoint.tex}}%
 \immediate\closeout\psbj@inaux
 \expandafter\joinfiles\GlobalInputList\into{psbjoint.tex}%
}%
%
%
%
\def\centinsert#1{\midinsert\line{\hss#1\hss}\endinsert}%
\def\psannotate#1#2{\vbox{%
  \def\ps@nnotation{#2\global\let\ps@nnotation=\relax}#1}}%
\def\pscaption#1#2{\vbox{%
   \setbox\drawingBox=#1
   \copy\drawingBox
   \vskip\baselineskip
   \vbox{\hsize=\wd\drawingBox\setbox0=\hbox{#2}%
     \ifdim\wd0>\hsize
       \noindent\unhbox0\tolerance=5000
    \else\centerline{\box0}%
    \fi
}}}%
%
\def\at(#1;#2)#3{\setbox0=\hbox{#3}\ht0=0pt\dp0=0pt
  \rlap{\kern#1\vbox to0pt{\kern-#2\box0\vss}}}%
%
\newdimen\gridht \newdimen\gridwd
\def\gridfill(#1;#2){%
  \setbox0=\hbox to 1\pscm
  {\vrule height1\pscm width.4pt\leaders\hrule\hfill}%
  \gridht=#1
  \divide\gridht by \ht0
  \multiply\gridht by \ht0
  \gridwd=#2
  \divide\gridwd by \wd0
  \multiply\gridwd by \wd0
  \advance \gridwd by \wd0
  \vbox to \gridht{\leaders\hbox to\gridwd{\leaders\box0\hfill}\vfill}}%
%
\def\fillinggrid{\at(0cm;0cm){\vbox{%
  \gridfill(\drawinght;\drawingwd)}}}%
%
%
\def\textleftof#1:{%
  \setbox1=#1
  \setbox0=\vbox\bgroup
    \advance\hsize by -\wd1 \advance\hsize by -2em}%
\def\textrightof#1:{%
  \setbox0=#1
  \setbox1=\vbox\bgroup
    \advance\hsize by -\wd0 \advance\hsize by -2em}%
\def\endtext{%
  \egroup
  \hbox to \hsize{\valign{\vfil##\vfil\cr%
\box0\cr%
\noalign{\hss}\box1\cr}}}%
%
\def\frameit#1#2#3{\hbox{\vrule width#1\vbox{%
  \hrule height#1\vskip#2\hbox{\hskip#2\vbox{#3}\hskip#2}%
        \vskip#2\hrule height#1}\vrule width#1}}%
\def\boxit#1{\frameit{0.4pt}{0pt}{#1}}%
\catcode`\@=12 
%
\psfordvips   

%% file: prd2000.bbl
\begin{thebibliography}{99}
\bibitem{L75}
   V.~N.~Gribov and L.~N.~Lipatov, Sov.~J.~Nucl.~Phys. {\bf 15} (1972) 438; 675;
              L.~N.~Lipatov, Sov.~J.~Nucl.~Phys. {\bf 20} (1975) 94;
              Y.~L.~Dokshitser, JETP ~{\bf 46} (1977) 641;
              G.~Altarelli and G.~Parisi, Nucl.~Phys. {\bf 126} (1977) 298.
\bibitem{BL80} S.~J.~Brodsky, and G.~P.~Lepage,  Phys. Lett. {\bf B87} (1979)
359; Phys.~Rev. {\bf D22} (1980) 2157;
A.~V.~Efremov and A.~V.~Radyushkin,  Phys.Lett~{\bf B94} (1980) 245;
Theor. Math. Phys. {\bf 42} (1980) 97.

\bibitem{CFP80} G.~Curci, W.~Furmansky, R.~Petronzio, Nucl.~Phys. {\bf B175}
 (1980) 27.
\bibitem{FLK81} E.~G.~Floratos, R.~Lacaze and C.~Kounnas,  Phys. Lett.
~{\bf B 98} (1981) 89; 285.
\bibitem{DR84} F.~M.~Dittes and A.~V.~Radyushkin, Phys.~Lett. {\bf B134} (1984) 359;
M.H. Sarmadi, Phys.~Lett. {\bf B143} (1984) 471;  ~S.V. Mikhailov
and A.V.Radyushkin, {\em ``Evolution kernel for the pion wave
function: two loop QCD calculation in Feynman gauge."}. Dubna
preprint JINR P2-83-721-mc (1983).
\bibitem{MR85} S.~V.~Mikhailov and A.~V.~Radyushkin, Nucl.~Phys. {\bf B254}
 (1985) 89.
\bibitem{LRV94}
 S.~A.~Larin, T.~van~Ritbergen, J.~A.~M.~Vermaseren, Nucl.~Phys. {\bf B427}
 (1994) 41;\\
 S.~A.~Larin, P.~Nogueira, T.~van~Ritbergen, J.~A.~M.~Vermaseren,
 Nucl.~Phys. {\bf B492} (1997) 338.
\bibitem{M97} S.~V.~Mikhailov, Phys.~Lett. {\bf B416} (1998) 421.
\bibitem{VPH81} A.~N.~Vasil'ev, Yu.~M.~Pis'mak and J.~R.~Honkonen,
 Theor.~Math.~Phys. 46 (1981) 157;  47 (1981) 291;
 A.~N.~Vasil'ev and M.~Yu.~Nalimov,
 Theor.~Math.~Phys. 55 (1982) 163;  56 (1983) 15.
\bibitem{Gr94}
J.~A.~Gracey,  Phys.~Lett. {\bf B322} (1994) 141;
Nucl. Phys. ~{\bf B480} (1996) 73.
\bibitem{P-M-P84} A.~Palanques-Mestre and P.~Pascual, Comm.~Math.~Phys.
95 (1984) 277; M.~Beneke and V.~M.~Braun, Nucl.~Phys. {\bf B426}
(1994) 301.
\bibitem{MS98} S.~V.~Mikhailov, Phys.~Lett. {\bf B431} (1998) 387.
\bibitem{BrGr95}
D.~J.~Broadhurst and A.~G.~Grozin, Phys.~Rev. {\bf D}52 (1995)
4082.
\bibitem{GK97} P.~Gosdzinsky and N.~Kivel, Nucl.~Phys. {\bf B521} (1998) 274,
 hep-ph/9707367.
\bibitem{IZ} C.~Itzykson and J-B.~Zuber, Quantum field theory
(Mc Graw-Hill. Inc.,1995), Chapter 12.
\bibitem{Ch96}
K.~G.~Chetyrkin, A.~H.~Hoang, J.H.~Kuhn, M.~Steinhauser, T.~Teubner,
 Phys.~Lett. {\bf B384} (1996) 233.
\bibitem{sirlin95}
P.~Gambino, A.~Sirlin,  Phys.~Lett. {\bf B355} (1995) 295;
K.~Philippides, A.~Sirlin,  Nucl.~Phys. {\bf B450} (1995) 3;
\bibitem{field98}
J.~H.~Field, hep-ph/9811399 (unpublished).
\bibitem{MMS97} L.~Mankiewicz, M.~Maul and E.~Stein,
Phys.~Lett {\bf 404} (1997) 345.
\bibitem{BLM83} S.~J.~Brodsky, G.~P.~Lepage  and B.~Mackenzie,  Phys.~Rev. {\bf D28} (1983) 228;
\bibitem{MR86} S.~V.~Mikhailov and A.~V.~Radyushkin, Nucl.~Phys. {\bf B273} (1986) 297.
\bibitem{bv96}
R.~Kirschner and L.~N.~Lipatov, Nucl.~Phys. {\bf 213} (1983)
122;\\ J.~Blumlein and A.~Vogt,  Phys.~Lett. {\bf B370} (1996)
149.
\bibitem{CDGM99}
M.~Ciuchini, S.~E.~Derkachov, J.~A.~Gracey and A.~N.~Manashov,
 Phys.~Lett. {\bf B458} (1999) 117.
\bibitem{LV93}
 S.~A.~Larin, J.~A.~M.~Vermaseren, Phys.~Lett {\bf 303} (1993) 334.
 \bibitem{BE} H.~Bateman and A.~Erdelyi, {\it Higher Transcendental Functions},
 (McGraw-Hill, 1953) Vol. 1
\bibitem{Mu95}D.~M\"uller,
Phys.~Rev. {\bf D 51} (1995) 3855.
\bibitem{BeMul97}A.~V.~Belitsky and D.~M\"uller, Phys.~Lett. {\bf B417} (1998) 129.
\bibitem{Gr96}
J.~A.~Gracey,  Phys.~Lett. {\bf B373} (1996) 178; Nucl. Ins. Meth.
Res. ~{\bf A389} (1997) 361.
\end{thebibliography}
